# ALIGNMENT OF THE ATLAS INNER DETECTOR

M. Ahsan (On behalf of the ATLAS Collaboration)
*University of Texas at Dallas, Richardson, TX 75080, USA*

ATLAS is one of the two general purpose detectors at the world's largest particle accelerator, the Large Hadron Collider (LHC). The LHC will be colliding proton beams at a center of mass energy $\sqrt{s}$= 14 TeV and is currently operating at $\sqrt{s}$ = 7 TeV. During the commissioning phase since September 2008, the ATLAS recorded cosmic-rays data and proton-proton collisions at $\sqrt{s}$ = 0.9 TeV. This data has extensively been used for the alignment and calibration of various sub-detectors. The ATLAS detector has a precision tracking system installed around the beam pipe for the measurements of the position and momentum of charged particles emerging from the collisions. The precise knowledge of misalignments of the tracking devices is crucial for the important physics measurements. At the time of writing the alignment corrections were obtained from the cosmic-rays and 0.9 TeV proton-proton collisions data, while the large statistics of proton-proton collisions at 7 TeV was used to check the performance of the alignment. This article gives an overview of the alignment strategy and the alignment performance using the data collected from proton-proton collisions at 7 TeV.

## 1. INTRODUCTION

The ATLAS inner detector (ID), as shown in Fig. 1, consists of position devices near the interaction point. It comprises three sub-systems based on silicon and drift-tubes technologies. These are called the Pixel detector (Pixels), the semiconductor tracker (SCT), and the transition radiation tracker (TRT) [1]. The Pixels, closest to the beam pipe, has three barrel layers and three end-cap disks on either sides of the barrel. It provides an intrinsic spatial resolution of 14 μm × 115 μm in the $r\varphi \times z$ direction. The SCT consists of four barrel layers and nine end-cap disks on either sides of the barrel. An SCT module is composed of two sets of silicon strip sensors of 80 μm pitch laid back-to-back with 40 mrad stereo angles, providing spatial resolution of 17 μm in the $r\varphi$ direction. The outermost ID tracking system, the TRT, is a gaseous system of straw tubes with anode wires, where straws are arranged in three cylindrical layers of 96 barrel modules and 14 disks on either side of the TRT barrels. An intrinsic spatial resolution is 130 μm in the $r\varphi$ direction for the TRT modules. The ID system is surrounded by superconducting solenoid providing a 2 T magnetic field.

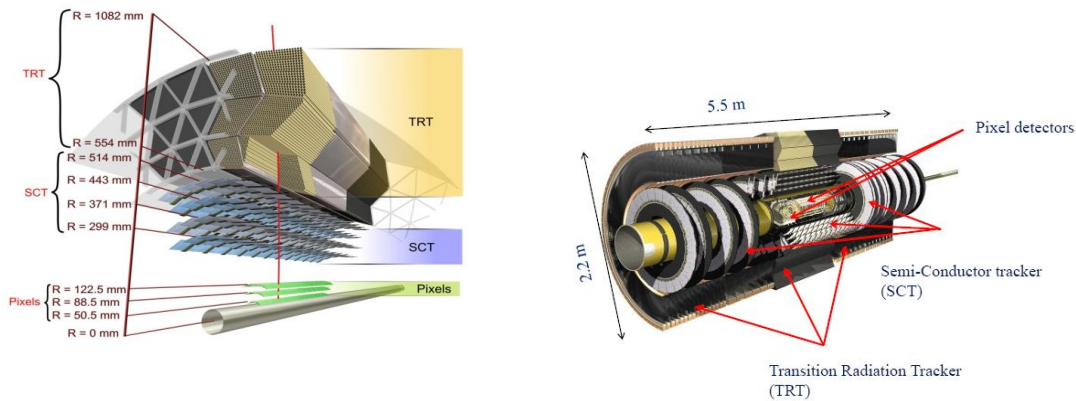

Figure 1: Schematic view of the ATLAS Inner Detector.

Alignment of the precision tracking devices is very important for physics measurements. In order not to degrade the track parameter resolution by more than 20%, the alignment precision should be up to 7 μm for the Pixels, 12 μm for the SCT, and 30 µm for the TRT. There are above six thousand alignable modules in the ID system which can have movements in up to six degrees of freedom (DoFs), three translational and three rotational. Therefore, the alignment of a large tracking system requires development of sophisticated techniques.

## 2. INNER DETECTOR ALIGNMENT STRATEGY

In order to perform quasi-real time position measurements of the tracking devices, the SCT is equipped with frequency scanning interferometry (FSI) [2], an optical alignment monitoring system which is currently in commissioning phase. A procedure complementary to hardware-based alignment is employed to perform absolute position measurements using the ID reconstructed tracks at ATLAS. This procedure is called track-based alignment and is discussed below.

### 2.1. Track-based alignment

The basic principle of track-based alignment is the minimization of $\chi^2$ formed from track-hit residual vectors, $r(a,\pi)$, where $a$ and $\pi$ denote the alignment parameter and the track parameter, respectively.

$$\chi^2 = \sum_{tracks} r^T V^{-1} r,$$

where $r(a,\pi)$ is defined as the distance between the measured hit position and the extrapolated track intersection in the module plane, and $V$ is the covariance matrix of hit measurements. The summation runs over all selected tracks in many events. The alignment corrections are obtained by minimizing $\chi^2$ with respect to the alignment parameters $a$. Several alignment algorithms [3-6] developed at ATLAS utilize track-hit residuals.

As the size of misalignments in general has some dependence on the size of the alignable structures, the alignment is performed at different levels of granularity. The scale of the problem changes from 24 DoFs to 36000 DoFs as the alignment granularity level increases. The larger structures, e.g the Pixel detector as a whole, the SCT barrels and the SCT end-caps, are aligned first, as it helps correcting the collective misalignment of the large structures.

Alignment algorithms are operated iteratively on the reconstructed and refitted tracks until the shifts in alignment parameters are converged to nearly zero.

### 2.2. Weak modes

Global deformations present in the detector that would leave the track-hit residuals and track $\chi^2$/DoF unchanged are called weak modes. Such weak modes produce biases in the reconstructed track parameters. Some of these systematic misalignments which affect physics the most are studied using the Monte Carlo simulation [7]. The cosmic-ray and beam halo data can serve as a good tool to tackle these weak modes. Using the limited data statistics no significant indication of weak modes has been found as of writing of this proceeding.

## 3. ALIGNMENT PERFORMANCE USINIG LHC COLLISION DATA

Until the start of data taking with proton-proton ($pp$) collisions at the LHC, the ATLAS has been collecting the cosmic-ray data with different magnetic field configurations: solenoid on and off. The tracks collected from cosmic-rays were used to produce the first set of ID alignment constants. These tracks give non-uniform illuminations to the detector and have different incident angles to the module planes in the barrel and the end-cap regions. After the LHC start-up in November 2009, the cosmic-ray tracks were combined with the tracks collected from $pp$ collisions at 0.9 TeV to perform the alignment. This significantly improved the alignment, particularly in the end-caps.

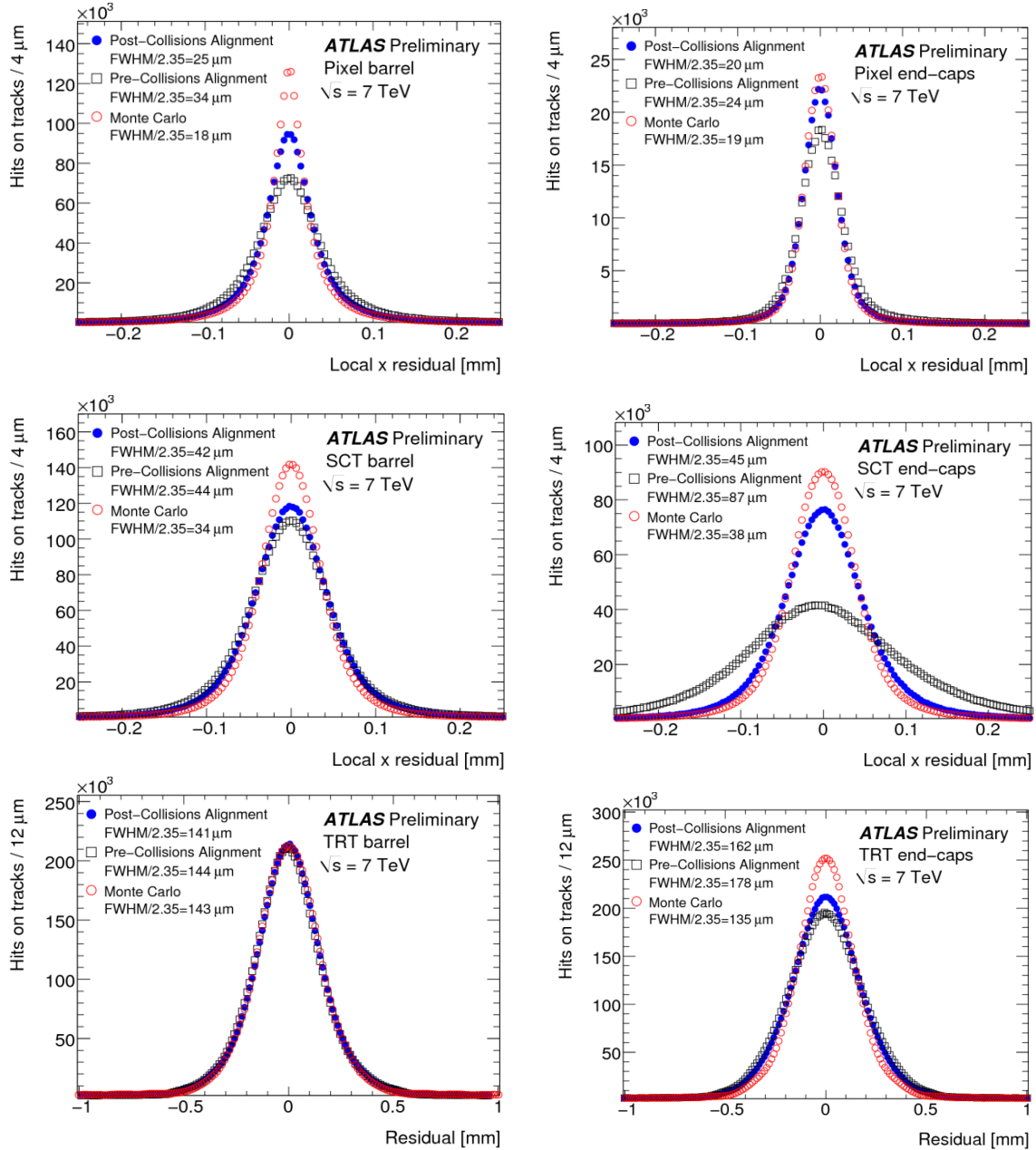

Figure 2: Track-hit residual distributions for barrel and end-caps in the Pixels, the SCT and the TRT. The tracks are required to have a minimum transverse momentum $p_T$ of 10 GeV.

As of writing the alignment corrections were obtained from the tracks collected from cosmic-rays and $pp$ collisions at 0.9 TeV collisions, and the performance of the alignment was checked using the large statistics of tracks collected from $pp$ collisions at 7 TeV data as shown in Fig. 2 [8]. The alignment constants obtained before (after) using the tracks from collisions at 0.9 TeV is labeled as 'Pre-Collisions Alignment' ('Post-Collisions Alignment'). The widths of the residual distributions are approaching those of the simulation with perfect knowledge of geometry, and the consistent performance have been found on both the 0.9 TeV and 7 TeV collisions indicating very good stability of the detector. The alignment is currently being worked on using high statistics of 7 TeV data to further improve the performance.

## 4. CONCLUSIONS

The track-based alignment procedure has been utilized to align the ATLAS Inner Detector using the tracks collected from cosmic-rays and $pp$ collisions at 0.9 TeV. The widths of the track-hit residual distributions from collision data are approaching those of the simulation. Further improvement is expected using large statistics of high momentum tracks from $pp$ collisions at 7 TeV.

## References

[1]   G. Aad et al. [The ATLAS Collaboration], The ATLAS Experiment at the CERN Large Hadron Collider, JINST3 (2008) S08003.

[2]   P.A. Coe, D.F. Howell, R.B. Nickerson, Meas. Sci. Technol. **15** (2004) 2175-2187; S. Gibson et al., Opt. Las. Eng. **43**, Issue 7 (2005) 815-831.

[3]   P. Buckmann de Renstorm, A. Hicheur and S. Haywood, ATL-INDET-PUB-2005-002 (2005).

[4]   R. Härtel, Diploma thesis, TU München, Germany, (2005).

[5]   T. Göttfert, Diploma thesis, Universität Würzburg, MPP-2006-118 (2006).

[6]   F. Heinemann, ATL-INDET-PUB-2007-011 (2007).

[7]   G. Aad et al. [The ATLAS Collaboration], ATL-PHYS-PUB-2009-080 (2009).

[8]   G. Aad et al.,[The ATLAS Collaboration], ATLAS-CONF-2010-067 (2010).